\documentclass{epl}
\usepackage{epsf}

\newcommand{\be}{\begin{equation}}
\newcommand{\ee}{\end{equation}}
\newcommand{\bea}{\begin{eqnarray}}
\newcommand{\eea}{\end{eqnarray}}

\shorttitle{Quantum-critical superconductivity\ldots}
\shortauthor{Ar. Abanov \etal}
\institute{
     \inst{1} Department of Physics, University of Wisconsin, 
Madison, WI 53706\\   
     \inst{2} Department of Physics and Ames Laboratory, 
 Iowa State University, Ames, IA 50011}   
 

\begin{document}

\title{Quantum-critical superconductivity in underdoped cuprates}
\author{Ar. Abanov\inst{1} \and Andrey V. Chubukov\inst{1} \and J\"org Schmalian%
\inst{2}}
\maketitle

\begin{abstract}
We argue that the pseudogap phase may be
 an attribute of the non-BCS pairing of
quantum-critical, diffusive fermions 
near the antiferromagnetic quantum critical point.
We derive and solve a set of three coupled Eliashberg-type equations for
spin-mediated pairing and show that in some $T$ range 
below the pairing instability, there is no feedback from superconductivity
on fermionic excitations, and
fermions remain diffusive despite of the pairing. We conject 
that in this regime, fluctuations of the pairing gap destroy the
 superconducting condensate but preserve the leading edge gap in the fermionic spectral function. 
\end{abstract}

The pseudogap behavior in underdoped cuprates is one of the most unusual and
exciting features of condensed matter physics. By all experimental accounts,
below optimal doping superconducting-like behavior of fermionic observables 
sets in at a temperature
which increases with underdoping. This temperature correlates with the value
of the superconducting gap at $T=0$, but does not correlate with the
transition temperature itself. The latter decreases with underdoping and
eventually vanishes\cite{timusk}.

In this paper we argue that the pseudogap behavior may be a part of new,
non-Fermi-liquid physics associated with the pairing of fermions in the
quantum-critical regime near the antiferromagnetic quantum critical point.
The onset of the pairing in this regime has been recently studied by
Finkel'stein and two of us (ACF)~\cite{acf}. ACF demonstrated that the onset
temperature $T = T_{\mathrm{ins}}$ is determined by a competition between
fermionic incoherence and the absence of a gap for spin excitations which
mediate superconductivity\cite{pines}. Due to this competition, $T_{\mathrm{%
ins}}$ tends to a finite value when the spin correlation length, $\xi $,
diverges.

An obvious question posed by this result is whether the pairing instability
at $T_{\mathrm{ins}}$ leads to true superconductivity, or only gives rise to
a formation of spin singlets which still behave incoherently and hence do
not superconduct. If this was the case, the phase right below $T_{\mathrm{ins}%
} $ would be a pseudogap phase, while the actual superconductivity would
emerge only at a smaller temperature.

This paper is a first step in addressing this issue. We report the results
of our analysis of the system behavior within the Eliashberg-type approach~ 
\cite{eliash}. This approach is mean-field like in the sense that it does
not include the feedback effect from the pairing modes (e.g., phase
fluctuations) on fermions and spin excitations. However, it does include
nontrivial physics associated with non BCS pairing in the quantum-critical
regime. We show that pairing of quantum critical fermions causes a reduction
of the superfluid stiffness, $D_{s}\propto n_{s}/m^{\ast }$. At low $T$, the
reduction is predominantly due to an enhancement of the quasiparticle mass, $%
m^{\ast }$. However, in some temperature interval between $T_{ins}$ and $%
T_{0}\leq T_{ins}$ we found an additional reduction of the superfluid
stiffness. We argue that the latter effect is an indication of a new physics
associated with a quantum-critical, non-BCS pairing near the magnetic
instability, and reflects the reduction of the superfluid density $n_{s}$
due to the fact that fermions remain diffusive at the lowest $\omega $
despite the formation of pairs. 
We conject that phase fluctuations, acting on top of the Eliashberg
solution, likely destroy fermionic coherence at $T\sim T_{0}$, which then
becomes a true phase transition temperature where a continuous $U(1)$
symmetry breaks down. The pairing gap, defined as the scale below which the
quasiparticle spectral weight is reduced, however, remains finite at $T_{0}$
and disappears only at $T_{ins}$. 

We now proceed with the calculations. Our theoretical treatment is based on
the spin-fermion model which describes the interaction between low-energy
fermions and their collective spin degrees of freedom peaked at 
 $\mathbf{Q}=(\pi
,\pi )$~\cite{acf,ac}. The
model has two energy scales: the effective spin-fermion interaction, $\bar{g}%
=g^{2}\chi _{0}$, where $g$ is the spin-fermion coupling, and $\chi _{0}$ is
the overall factor in the static spin susceptibility $\chi _{q}=\chi _{0}/(\xi ^{-2}+(\mathbf{q}-\mathbf{Q})^{2})$, and
a typical fermionic energy, $v_{\mathrm{F}}\xi ^{-1}$, where $v_{\mathrm{F}}$
is the Fermi velocity. The dimensionless ratio of the two, $\lambda =3{\bar{g%
}}/(4\pi v_{\mathrm{F}}\xi ^{-1})$, determines the relative strength of the
spin-fermion coupling. 
The quantum-critical behavior obviously corresponds to the strong coupling
limit, $\lambda \geq 1$ 
~\cite{ac}.

For superconductors with  phonon mediated pairing, a recipe to study the
system behavior at strong coupling is the Eliashberg theory. It is justified
by Migdal theorem which states that the corrections to the electron-phonon
vertex $\delta g/g$ and $v_{F}^{-1}d\Sigma (k,\omega )/dk$, although
increase with the dimensionless coupling $\lambda $, still scale as $\lambda
(v_{s}/v_{F})$ where $v_{s}$ and $v_{F}\gg v_{s}$ are the sound velocity and
the Fermi velocity. As $v_{s}/v_{F}\sim 10^{-4}$, $\delta g/g$ and $%
v_{F}^{-1}d\Sigma (k,\omega )/dk$ can be safely neglected for all reasonable 
$\lambda $. One then has to include only $\Sigma (\omega )$~\cite{eliash}.
 In our case,
Migdal theorem is inapplicable as spin fluctuations are collective modes of
fermions and hence the spin velocity (the analog of $v_{s}$) is of the same
order as $v_{F}$. This implies that the Migdal parameter $\lambda v_{s}/v_{F}
$ is large in the strong coupling regime. It turns out, however, that in
this limit 
one
again can neglect $\delta g/g$ and $v_{F}^{-1}d\Sigma (k,\omega )/dk$, this
time because the dynamics of the collective mode is completely modified by
low-energy fermions and becomes  diffusive at energies relevant to
the pairing problem. [For electron-phonon interaction this happens only for
very small frequencies $\simeq $ $(v_{s}/v_{F})^{2}T_{c}\ll T_{c}$.]
Specifically, the strong coupling solution of the spin-fermion problem in
the normal state yields $d\Sigma /d\omega \propto \lambda $ while $\delta g/g
$ and $v_{F}^{-1}d\Sigma (k,\omega )/dk$ scale as $(1/N)\log \lambda \ll
\lambda $, where $N(=8)$ is the is the number of hot spots in the Brillouin
zone ~\cite{ac}. Furthermore, below $T_{ins}$, we found that $\delta g/g$
even becomes independent on $\lambda $. Physically, the irrelevance of
vertex corrections is due to the fact that the theory possesses no SDW
precursors, and hence spin excitations are not near Goldstone bosons~\cite
{vertex}. Below we formally treat $N$ as a large parameter and perform
computations at $N\rightarrow \infty $. In this limit, vertex corrections
can be totally neglected. Also, following ACF, we neglect the momentum
variation of $\Sigma _{\mathbf{k}}(\omega )$ and of the anomalous vertex $%
\Phi _{\mathbf{k}}(\omega )$ along the Fermi surface, i.e., replace them by $%
\Sigma _{\mathbf{k}_{\mathrm{hs}}}(\omega )$ and $\Phi _{\mathbf{k}_{\mathrm{%
hs}}}(\omega )$ (respecting the $d_{x^{2}-y^{2}}$ symmetry $\Phi _{\mathbf{k}%
_{\mathrm{hs}}+\mathbf{Q}}=-\Phi _{\mathbf{k}_{\mathrm{hs}}}$). The last
approximation is justified if ${\bar{g}}\ll v_{F}k_{F}$ which we assume to
hold. Finally, we verified that both above and below $T_{\mathrm{ins}}$,
there is no universal thermal contribution to $\xi $ from low-energy
fermions, i.e., $\xi (T)\simeq \xi (T=0)$. At $T>{\bar{g}}$, the physics is
dominated by classical fluctuations, and $\xi ^{-1}$ eventually becomes
strongly $T$ dependent~\cite{SPS}.

We next discuss the structure of the Eliashberg equations for spin-mediated
superconductivity. In conventional superconductors, the Eliashberg theory
involves two coupled equations for the fermionic self-energy $\Sigma $, and
the pairing vertex $\Phi $\cite{eliash}. Modifications of the phonon
propagator due to fermionic pairing are small and can be neglected. For
spin-mediated pairing, the situation is different as the spin dynamics is
made out of low-energy fermions and thus is sensitive to the opening of the
fermionic gap \cite{acf,ac}. As a result, the Eliashberg theory has to be
generalized to include the equation for the dynamical spin susceptibility.

The mutual feedbacks between fermions and spin fluctuations have been
earlier studied numerically within the FLEX approximation~\cite{sm,joerg}.
Our results 
agree with these studies, but go beyond them in the understanding of the new
physics. The modification of the spin propagator due to fermionic pairing
has also been considered in the context of marginal Fermi liquid
phenomenology~\cite{vl}, but the feedback effect on fermions has not been
studied.

The set of Eliashberg equations for the spin-fermion model is obtained in a
standard way, by evaluating diagrammatically the pairing vertex, fermionic
self-energy and the spin polarization operator, and integrating over the
fermionic energy. The results have been earlier presented in the two
limiting cases: for infinitesimally small $\Phi (\omega )$~ \cite{acf}, and
for $\Phi (\omega )=const$~\cite{ac}. A straightforward extension to a
general case yields ($\Phi (\omega _{m})=\Phi _{m}$, etc.), 
\begin{eqnarray}
\Phi _{m} &=&~\frac{\pi T}{2}\sum_{n}\frac{\Phi _{n}}{\sqrt{\Phi
_{n}^{2}+\Sigma _{n}^{2}}}~\left( \frac{{\bar{\omega}}}{\omega _{sf}+\Pi
_{n-m}}\right) ^{1/2},  \label{setphi} \\
\Sigma _{m} &=&\omega _{m}+\frac{\pi T}{2}\sum_{n}\frac{\Sigma _{n}}{\sqrt{%
\Phi _{n}^{2}+\Sigma _{n}^{2}}}~\left( \frac{{\bar{\omega}}}{\omega
_{sf}+\Pi _{n-m}}\right) ^{1/2},  \label{setsigma} \\
\Pi _{m} &=&\pi T~\sum_{n}\left( 1-\frac{\Sigma _{n}\Sigma _{n+m}+\Phi
_{n}\Phi _{n+m}}{\sqrt{\Phi _{n}^{2}+\Sigma _{n}^{2}}~\sqrt{\Phi
_{n+m}^{2}+\Sigma _{n+m}^{2}}}\right) .  \label{setpi}
\end{eqnarray}
The dynamical spin susceptibility at antiferromagnetic momentum $\mathbf{Q}$
and the fermionic Green's function follow as $\chi _{\mathbf{Q}}(\omega
_{m})^{-1}\propto \omega _{\mathrm{sf}}-\Pi _{m}$, and $G_{\mathbf{k}%
}(\omega _{m})=-(i\Sigma _{m}+\epsilon _{\mathbf{k}})/(\Phi _{m}^{2}+\Sigma
_{m}^{2}+\epsilon _{\mathbf{k}}^{2})$, respectively, where $\epsilon _{%
\mathbf{k}}=\mathbf{v}_{F}\cdot (\mathbf{k}-\mathbf{k}_{F})$. For further
convenience we included the bare $\omega _{n}$ term in the fermionic
propagator into $\Sigma _{n}$ and introduced ${\bar{\omega}}=9{\bar{g}}%
/(2\pi N)$ and $\omega _{\mathrm{sf}}={\bar{\omega}}/(4\lambda ^{2})\propto
\xi ^{-2}$ instead of $\lambda $ and ${\bar{g}}$ (the $1/N$ factor in ${\bar{%
\omega}}$ is eliminated after appropriate rescaling ${\bar{g}\rightarrow 
\bar{g}N}$ and $v_{\mathrm{F}}\rightarrow v_{\mathrm{F}}N$).

Analyzing $\Sigma $ and $\chi _{\mathbf{Q}}$ in the normal state, we find
that $\omega _{\mathrm{sf}}$ separates the Fermi liquid behavior at $\omega
<\omega _{\mathrm{sf}}$ from the quantum-critical behavior which holds
between $\omega _{\mathrm{sf}}$ and ${\bar{\omega}}\gg \omega _{\mathrm{sf}}$%
. Which of the two energies determines the superconducting properties of the
system? The onset of pairing was determined by ACF by linearizing with
respect to $\Phi $. They found that $\Phi $ emerges at a temperature $T_{%
\mathrm{ins}}$ which depends weakly on $\omega _{\mathrm{sf}}$ and saturates
at $T_{\mathrm{ins}}\approx 0.17{\bar{\omega}}$ for $\xi \rightarrow \infty $%
.\cite{acf} Thus, the onset of pairing is produced by quantum-critical
fermions with ${\bar{\omega}}\geq \omega \gg \omega _{\mathrm{sf}}$. We
consider the solution of the full set of nonlinear equations (\ref{setphi}-%
\ref{setpi}) we analyze whether there are any new crossover scales  below $%
T_{ins}$. 
\begin{figure}[tbp]
\centerline{
\epsfxsize=0.9\columnwidth
\epsfysize=1.6in
\epsffile{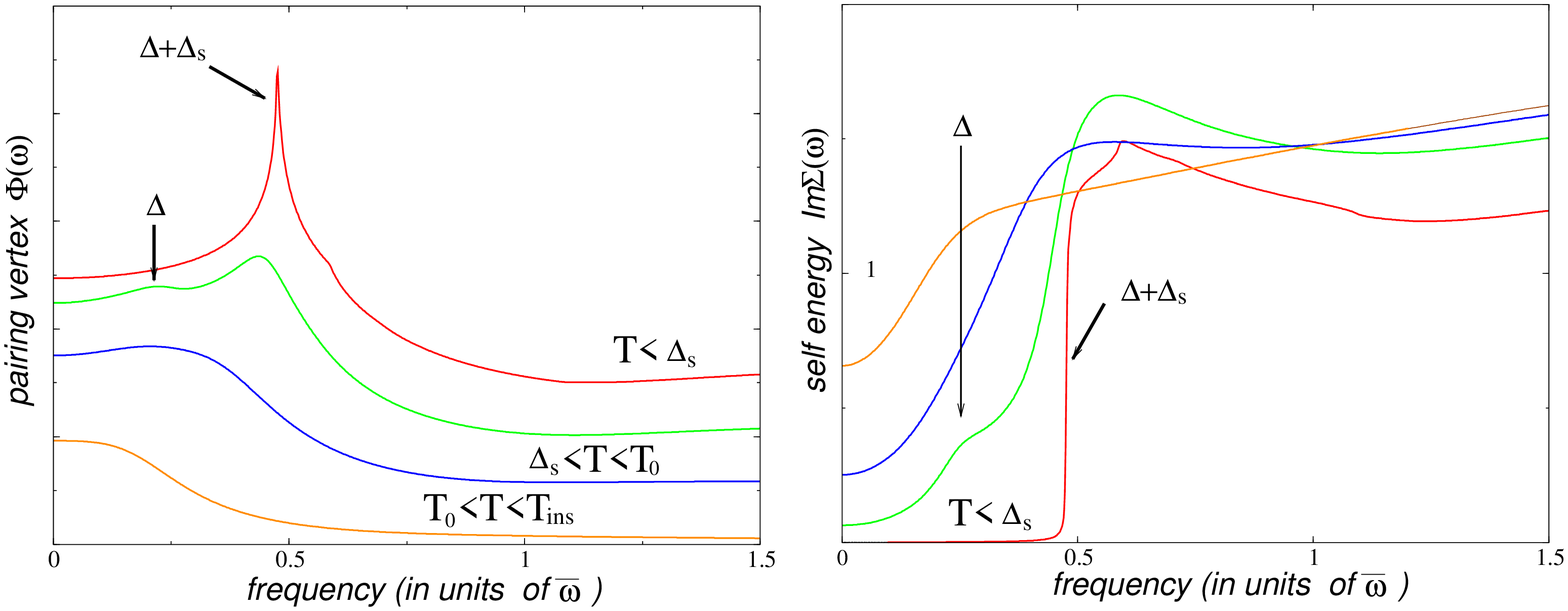}}
\caption{The real part of the pairing vertex $\Phi (\protect\omega )$ and
$Im \Sigma (\omega)$ for $%
\protect\lambda =4$ and different temperatures $T<\Delta _{s}$, $T\sim
\Delta _{s}$, $\Delta _{s}<T<T_{0}$ and $T_{0}<T<T_{ins}$. We associate $%
T\sim \Delta _{s}$ and $T\sim T_{0}$ with the onset of impurity-like
behavior due to thermal fluctuations, and with the onset of the reduction of
the superfluid density, respectively. }
\label{fig1}
\end{figure}
\begin{figure}[tbp]
\centerline{
\epsfxsize=0.9\columnwidth
\epsfysize=2.1in
\epsffile{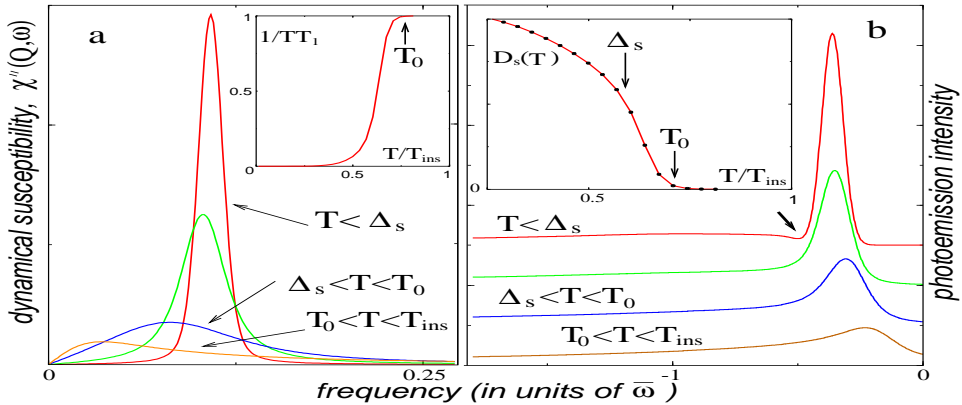}}
\caption{The dynamical spin susceptibility $\protect\chi ^{\prime \prime }(%
\mathbf{Q},\protect\omega )$ and the photoemission intensity $I(\protect%
\omega )=A(\protect\omega )n_{F}(\protect\omega )$ for $\protect\lambda =4$
at different $T$. The insets show the the superfluid stiffness $D_{s}(T)$
and the NMR relaxation rate $1/T_{1}T$ vs $T/T_{\mathrm{ins}}$. The actual
behavior of $D_{s}(T)$ and $1/T_{1}T$ at low $T$ is smoother due to
contributions from nodal regions. The arrow indicates the position of the dip in $I(\omega)$.}
\label{fig2}
\end{figure}

At $T=0$, the solution of the Eliashberg set is in many respects similar to
the one for $\Phi (\omega )=const$~ \cite{ac}. The pairing gap $\Delta $
(defined in the usual way as $\Delta (\omega )=\Phi (\omega )\omega /\Sigma
(\omega )$ and $\Delta (\omega =\Delta )=\omega $) scales with $T_{\mathrm{%
ins}}$ up to $\log \lambda $ corrections which, however, remain numerically
small even for $\lambda =20$. The presence of the gap eliminates
spin-fermion scattering at energies $\mathcal{O}(\Delta )$ and restores
fermionic and bosonic coherence. In particular, spin excitations, which were
purely relaxational in the normal state 
become propagating below $2\Delta $ with dispersion $\Omega _{\mathbf{q}%
}=(\Delta _{s}^{2}+v_{F}(\mathbf{q}-\mathbf{Q})^{2})^{1/2}$, where $\Delta
_{s}\sim (\bar{\omega}\omega _{sf})^{1/2}\propto T_{ins}/\lambda $. This
effect changes the dynamical exponent to $z=1$ and yields the resonance in
the inelastic neutron scattering at $\omega =\Delta _{s}$. The spin
resonance in turn affects fermionic excitations: both, $\mathrm{Im}\Sigma
(\omega )$ and $\mathrm{Im}\Phi (\omega )$ vanish at low $\omega $ due to a
lack of a phase space for single particle decay and jump to finite values at 
$\omega =\Delta +\Delta _{s}$. This gives rise to a dip in the fermionic
spectral function \cite{ac,ding}. 

These analytical results fully agree with
our numerical solution (the lowest $T$ results in Figs~\ref{fig1} and \ref
{fig2}). We found $2\Delta /T_{\mathrm{ins}}\approx 4$ for $\lambda \gg 1$
with downturn deviation at $\lambda \sim 1$. To a reasonable accuracy, $%
\Delta _{s}\sim 0.35~{\bar{\omega}}/\lambda $. The spectral function has a
peak-dip-hump structure, and the peak-dip distance exactly equals $\Delta
_{s}$. The numbers which we obtain are also consistent with experiment. We use data
for $\mathrm{Bi}2212$ for comparisons. Near optimal doping, which we
identify with $\lambda \sim 1$, we obtain $\Delta _{s}/{\bar{\omega}}\sim
0.25-0.3$ and $\Delta /{\bar{\omega}}\approx 0.2$. The value of ${\bar{\omega%
}} $ can be extracted from the photoemission data as a frequency where
nonlinear corrections from $Re\Sigma (\omega )$ to the quasiparticle
dispersion become irrelevant. This yields~ \cite{johnson} ${\bar{\omega}}%
\sim 150-160\mathrm{meV}$. Using this ${\bar{\omega}}$, we obtain $%
\Delta_s\sim 38-48\mathrm{meV}$ and $\Delta \sim 30-32\mathrm{meV}$ which
are in reasonable agreement with the neutron scattering\cite{keimer},
photoemission~\cite{photo} and tunneling~ \cite{tunn} data.

We now turn to $T>0$. A simple analysis of Eqs. (\ref{setphi},\ref{setsigma}%
) shows that classical thermal spin fluctuations (the ones with zero
Matsubara frequency) critically depend on $\omega _{\mathrm{sf}}$ and are
therefore relevant already at low $T$. These fluctuations account for
scattering with zero energy transfer and therefore act for spin-mediated $d$%
-wave paring in the same way as nonmagnetic, elastic impurities in $s$-wave
superconductors. We then use the same strategy as for the impurity problem 
\cite{ag} and introduce ${\tilde{\Phi}}_{m}$ and ${\tilde{\Sigma}}_{m}$ via $%
\Phi _{m}={\tilde{\Phi}}_{m}\eta _{m}$, $\Sigma _{m}={\tilde{\Sigma}}%
_{m}\eta _{m}$, where 
\begin{equation}
\eta _{m}=1+\frac{\pi T\lambda }{\sqrt{({\tilde{\Phi}}_{m})^{2}+({\tilde{%
\Sigma}}_{m})^{2}}}  \label{tr}
\end{equation}

Substituting $\Phi _{m}$ and $\Sigma _{m}$ into the Eliashberg set we obtain
after some algebra that the \emph{quantum contributions }to the self energy, 
${\tilde{\Phi}}_{m}$ and ${\tilde{\Sigma}}_{m}$, obey the same Eqs. (\ref
{setphi}), (\ref{setsigma}) but without zero frequency\textit{\ (}$m=n$)%
\textit{\ }contributions. This implies that the effects of thermal spin
fluctuations are completely absorbed into the $\eta _{m}$ factors. On the
other hand, the temperature dependence of ${\tilde{\Phi}}$ and ${\tilde{%
\Sigma}}$ is set by $T_{\mathrm{ins}}$ and is weak at $T\sim \Delta _{s}$.
Estimating $\eta $ 
using ${\tilde{\Sigma}}^{2}(\omega )+{\tilde{\Phi}}^{2}(\omega )\sim \Delta
^{2}$, we find $\eta -1\sim T/\Delta _{s}$. We see that at low $T\ll T_{ins}$
the system behaves as a  dirty superconductor, the role of $\gamma /\Delta $
ratio ( where $\gamma $ is the elastic scattering rate due to impurities) is
played by $T/\Delta _{s}$. Using this analogy and the results for dirty
superconductors~\cite{ag,Anderson} we find that the density of states and
the two-particle response functions at finite momentum, such as $\Pi \left(
\omega \right) $, are unaffected by $T/\Delta _{s}$ ratio as they have the
same form in terms of ${\tilde{\Phi}}_{m}$ and ${\tilde{\Sigma}}_{m}$ as in
terms of $\Phi _{m}$ and $\Sigma _{m}$. On the other hand, the single
particle spectral function and the two particle response functions at zero
external momentum (i.e., the Meissner kernel and the superfluid stiffness $%
D_{s}$) scale as $\eta ^{-1}$ and are substantially reduced above $T\sim
\Delta _{s}$. As in dirty superconductors, this reduction can be absorbed
into the renormalization of the quasiparticle mass $m^{\ast }/m=d{\Sigma}/d\omega \sim \eta $ and is not associated with the reduction of the
superfluid density.

These features are present in our numerical solution of the Eliashberg
equations. In Fig.~\ref{fig1} we show representative results for $\mathrm{Re}%
\Phi (\omega )$ and $\mathrm{Im} \Sigma (\omega)$ 
for different $T$. We clearly see that  at $T\sim \Delta _{s}$,  sharp structures 
at $\omega =\Delta +\Delta _{s}$ transform into
 broader structures at $\omega
=\Delta $. The latter are due to the fact that in real
frequencies, $\eta (\omega )$ is peaked at $\omega =\Delta $, and the
amplitude of this peak increases with $T$
  In
Fig.~\ref{fig2} we show the behavior of the dynamical spin susceptibility
and the fermionic spectral function (multiplied by $n_{F}(\omega )$) for
different $T$. The insets show the superfluid stiffness, $D_{s}(T)$,
extracted from the computation of the Meissner kernel, and the NMR
relaxation rate $1/T_{1}T\propto \left. \Pi (\omega )/\omega \right|
_{\omega \rightarrow 0}$. We see that the residue of the peak in the
spectral function, and the superfluid stiffness sharply decrease above $%
\Delta _{s}$. On the other hand, $1/T_{1}T$ and the peak in $\chi _{\mathbf{Q%
}}^{\prime \prime }\left( \omega \right) $ are much less sensitive to the
ratio $T/\Delta _{s}$. We explicitly verified by analyzing larger $\lambda
=20$ 
that $1/T_{1}T$ does not change much at $T\sim \Delta _{s}$. 

To this end, we therefore find a crossover in the system behavior at $T\sim
\Delta _{s}$, similar to a crossover at $\gamma \sim \Delta $ in dirty
superconductors. The superfluid stiffness is reduced at $T>\Delta _{s}$ due
to mass renormalization. However, just as in dirty superconductors, this
reduction of the stiffness \textit{does not} give rise to a substantial
reduction to $T_{c}$ due to phase fluctuations simply because still $%
D_{s}\sim E_{F}(m/m^{\ast })\gg \Delta ,T_{c}$. This follows from the fact
that within Eliashberg theory $\Sigma (\omega \sim \Delta )\sim \eta \Delta
\ll E_{F}$~\cite{eliash}.  
\begin{figure}[tbp]
\centerline{
\epsfxsize=0.9\columnwidth
\epsfysize=1.7in 
\epsffile{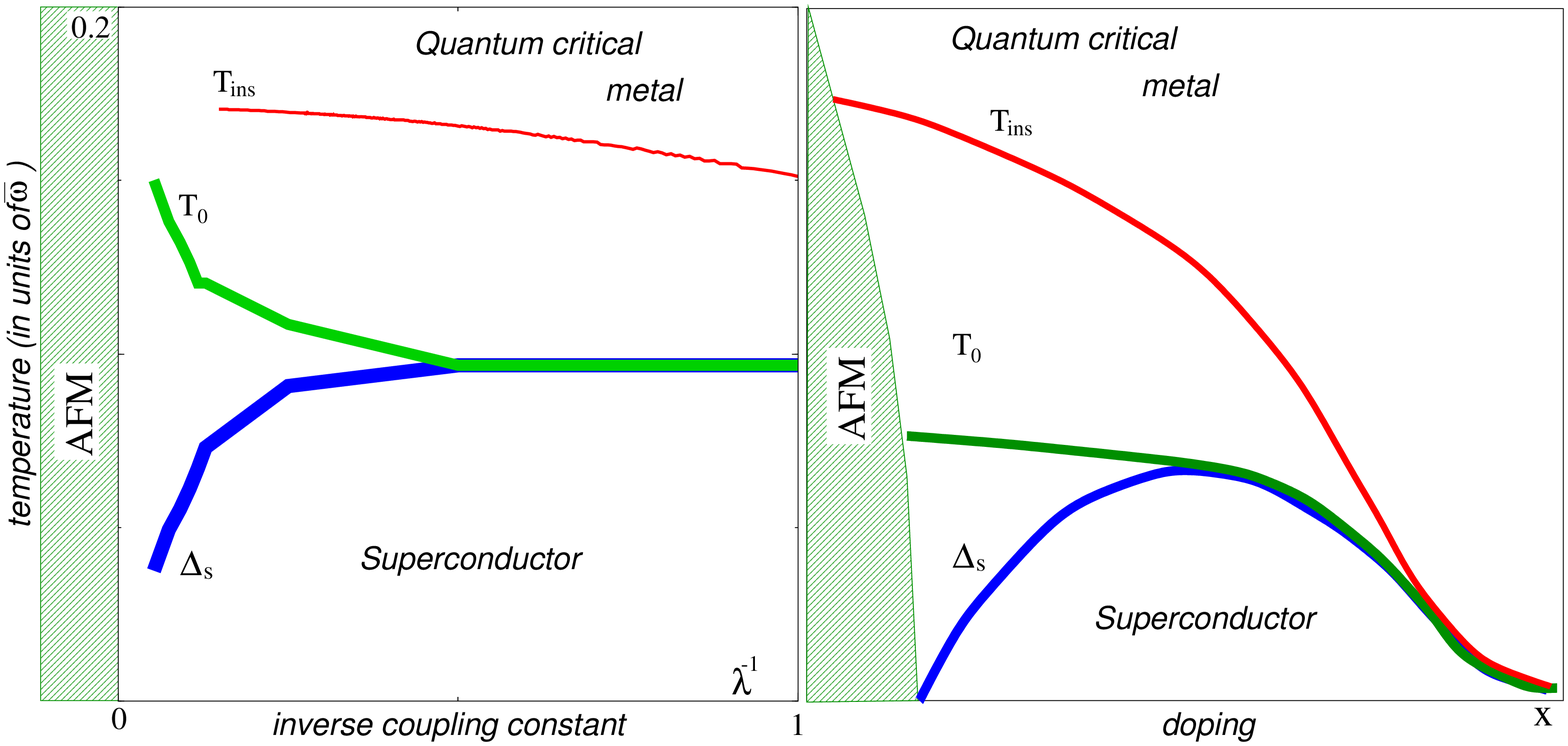}}
\caption{a). The phase diagram emerging from the solution of the
Eliashberg equations. Without phase fluctuations, $T=T_{\mathrm{ins}}$ is a
transition line, while the other two are the crossover lines. 
Above $\Delta_s$, superfluid stiffness is reduced due to enhancement of the
effective mass, like in dirty superconductors. 
Above $T_0$, it is further reduced due to a reduction
of the superfluid density. We conject that fluctuations destroy
superconductivity at $T \sim T_0 < T_{ins}$. b) The cuprate phase
diagram that follows from Fig. a).}
\label{fig3}
\end{figure}

We now argue that another physics emerges at $T\leq T_{ins}$ and yields an
extra reduction of $D_{s}$ unrelated to a mass renormalization. Indeed, we
clearly see in Fig~\ref{fig1} that the peak in $\mathrm{Re}\Phi (\omega )$
disappears at $T=T_{0}<T_{ins}$. Between $T_{0}$ and $T_{ins}$, $\mathrm{Re}%
\Phi (\omega )\neq 0$, but it monotonically decreases with frequency.
Simultaneously, 
 $\mathrm{Im}{\tilde{\Sigma}}(\omega)$ becomes roughly the
same as in the normal state, the peaks in the dynamical spin
susceptibility and in the fermionic spectral function become very broad,
 $1/T_{1}T$ nearly reaches its normal state value, and the
superfluid stiffness 
virtually disappears (see the insets in Fig.~\ref{fig2}). 
This
behavior is not associated with the impurity-like scattering of thermal
fluctuations as then the peak in $\mathrm{Re}\Phi (\omega )$ at $\omega
=\Delta $ would disappear only at $T_{ins}$. 
Rather it implies
that immediately below $T_{ins}$
there is no feedback from pairing on fermionic excitations in the
sense that strong $Im\Sigma (\omega =0)$ in the normal state does not convert
into a mass renormalization at $T<T_{ins}$. The reduction of $D_s$
without corresponding mass enhancement likely implies that the superfluid
density $n_{s}(T)$ is reduced compared to what it would be for a BCS $d-$%
wave superconductor at this temperature. We conject that
due to this extra strong reduction of $D_{s}$,
fluctuations of the gap destroy coherent fermionic pairing at $T\sim
T_{0} < T_{ins}$~\cite{ek}. 
Still, however, the leading edge gap in the spectral function remains finite
for all $T<T_{ins}$ and just fills in by fluctuations at $T\sim T_{0}$.

The phase diagram which emerges from our consideration  is presented in Fig.~%
\ref{fig3}. Without fluctuations, there is a true transition at $T=T_{%
\mathrm{ins}}$,  and two crossover lines at $T \sim T_{0}$ and $T \sim
\Delta_s$. For $\lambda \gg 1$, $\Delta_s\ll T_0\leq T_{\mathrm{ins}}$. We
found that $T_0$ and $\Delta_s$ merge at $\lambda \sim 2$, but still both
remain smaller than $T_{\mathrm{ins}}$. For smaller $\lambda $, the distance
between $T_0$ and $T_{\mathrm{ins}}$ gradually decreases and eventually
disappears. With fluctuations, it is likely that coherent superconductivity
appears only at $T\sim T_{0}$. We cannot argue definitely whether this
implies that $T_{ins}$ becomes a crossover temperature below which the
system begins creating disordered singlet pairs which condense at $T_{0}$,
or there is an Ising-like transition at $T\sim T_{ins}$ where singlets are
ordered into columnar dimers. The first possibility is, in our opinion, a
realization of Anderson's RVB idea~\cite{RVB}. The second possibility is a
realization of the scenario suggested by Vojta and Sachdev \cite{subir}.

We conclude the paper by summarizing what we obtained. Our key result is the
phase diagram in Fig.\ref{fig3}. We found that there are two different
fluctuation effects which govern the system behavior below the onset of the
pairing instability. First effect is due to thermal spin fluctuations, and
its role is equivalent to that of nonmagnetic impurities in $s-$wave
superconductors. This effect may account for the reduction of the superfluid
stiffness, but cannot destroy superconductivity. The second effect is a
quantum one 
 - we found that below $T_{ins}$, there
is no immediate feedback effect on fermionic self-energy,
and low-frequency fermionic excitations
remain overdamped despite pairing. In this situation, singlet pairs still
 diffuse rather than propagate, and superconducting condensate
 is easily destroyed by fluctuations.

The data for cuprates seem to indicate that $T_{c}$ scales with the
superfluid stiffness at $T=0$~\cite{oxford} and also with the resonance
neutron frequency $\Delta _{s}$~\cite{keimer}. To account for these results
in our theory, it is necessary that $T_{0}$ and $\Delta _{s}$ coincide. This
does happen at intermediate $\lambda $ (and our theory does predict that
near optimal doping $T_{c}$ scales with $\Delta _{s}$) but not at $\lambda
\gg 1$. In other words, the present theory  underestimates quantum
fluctuations at very strong coupling.

A final remark. Our results bear some similarities but also some discrepancies with the results of the Eliashberg study of
 phonon superconductors at vanishing Debye frequency~\cite{rolan}. 
The comparison with phonon case requires a separate study 
and will be presented elsewhere.
\acknowledgments
It is our pleasure to thank A. M. Finkel'stein for stimulating discussions
on all aspects of the paper. We are also thankful to P.W. Anderson, G.
Blumberg, J.C. Campuzano, S. Chakravarty, L.P. Gor`kov, L. Ioffe, P.
Johnson, R. Joynt, B. Keimer, D. Khveschenko, G. Kotliar, R. Laughlin, A.
Millis, D. Morr, M. Norman, D. Pines, S. Sachdev, D. Scalapino, Q. Si, and A.
Tsvelik for useful conversations. The research was supported by NSF
DMR-9979749 (Ar. A and A. Ch.), and by the Ames Laboratory, operated for the
U.S. DoE by ISU under contract No.
W-7405-Eng-82 (J.S).


\end{document}